\documentclass[universe,article,accept,moreauthors,pdftex]{Definitions/mdpi} 

\makeatletter
\g@addto@macro{\UrlBreaks}{\UrlOrds}
\makeatother

\usepackage{booktabs}

\firstpage{1} 
\pubvolume{xx}
\issuenum{1}
\articlenumber{5}
\pubyear{2020}
\copyrightyear{2020}
\history{Received: date; Accepted: date; Published: date}
\updates{yes} % If there is an update available, un-comment this line

\usepackage{amsmath} 
\usepackage{natbib}
\usepackage{pdflscape}
\Title{A Chi-Squared Analysis of the Measurements of Two Cosmological Parameters Over Time%Attention AE/ME. The following layout issues have not been checked by the English Editing Department and must be carefully verified by the AE/Layout Department: All callout issues, bold usage of callouts, and references to callouts in the text. Correct callout usage in figures. Figure and Table layout issues. Footnote formatting and Glossaries have not been checked. En dash usage for negative values, en dash usage to indicate relationships, en dash usage to indicate bonds (especially in chemistry). The English Editing Department is not responsible for correct italic usage for genes, proteins and technical terminology. This responsibility belongs to the authors. The following are also not checked: spacing between numbers and units of measurement, ratios, en dashes for ranges, date and time formats, punctuation in equation lines, and less than/more than spacing (< >). Finally, capitalization and layout of titles/headings must be properly checked as well as ensuring 'Eq.' and 'Fig.' are properly spelled out, as these are layout issues.
}

\Author{Timothy Faerber %Please carefully check the accuracy of names and affiliations. Changes will not be possible after proofreading.
 $^{1}$ and %MDPI: There is no explanation for ``†'' and ``‡'', please confirm and modify.
 Mart\'\i n L\'opez-Corredoira $^{2,3,}$*}

\AuthorNames{Timothy Faerber and Mart\'\i n L\'opez-Corredoira}

\address{%
$^{1}$ \quad Department of Physics and Astronomy, Uppsala University, P.O. Box 256, SE-751 05 Uppsala, Sweden; Timothy.Faerber.5858@student.uu.se\\
$^{2}$ \quad Instituto de Astrofisica de Canarias, E-38205 La Laguna, Spain\\
$^{3}$ \quad Departamento de Astrofisica, Universidad de La Laguna, E-38206 La Laguna, Spain
}

\corres{Correspondence: fuego.templado@gmail.com %This E-mail is different from Redmine. Please confirm.
}

\abstract{
The aim of this  analysis was to determine whether or not the given error bars truly represented the dispersion of values in a historical compilation of two cosmological parameters: the amplitude of mass 
   fluctuations ($\sigma_8$) and Hubble's constant ($H_0$) parameters in the standard cosmological model.
   For this analysis, a chi-squared test was executed on a compiled list of past measurements.
 It was found through  analysis of the chi-squared ($\chi^2$) values of the data that for $\sigma_8$ {(60 data points measured between 1993 and 2019
   and $\chi^2$ between 182.4 and 189.0)} the associated probability Q is extremely low, with $Q = 1.6 \times 10^{-15}$ for the weighted 
   average and $Q = 8.8 \times 10^{-15}$ for the best linear fit of the data. This was also the case for the $\chi^2$ values of $H_0$ 
   {(163 data points measured between 1976 and 2019 and $\chi^2$ between 480.1 and 575.7)}, where $Q = 1.8 \times 10^{-33}$ for the 
   linear fit of the data and $Q = 1.0 \times 10^{-47}$ for the weighted average of the data.
  The general conclusion was that the statistical 
   error bars associated with the observed parameter measurements have been underestimated or the systematic errors were not properly taken into account in at least 20\% of the measurements. {The~fact that the underestimation of error bars for $H_0$ is so common might explain the apparent 4.4$\sigma$ discrepancy formally known today as the Hubble tension.} 
}


\begin{document}
%%%%%%%%%%%%%%%%%%%%%%%%%%%%%%%%%%%%%%%%%%

\section{Introduction}
\label{S:1}
\unskip
\subsection{The Standard Cosmological~Model}

{The standard cosmological model is a model that aims to describe the evolution and structure of the Universe that we live in. This theoretical model accounts for our Universe's beginning through inflation caused by the Big Bang all the way up to the present-day dark energy dominated Universe ($\sim $70\%). In~addition to explaining the evolution and current state of the Universe, the~standard cosmological model can be interpreted to predict the Universe's fate. The~standard cosmological model consists of 12 parameters \citep{croft2015measurement}:} $\Omega_M$ is the ratio of the current matter density to the critical density, $\Omega_{\Lambda}$ is the cosmological constant as a fraction of the critical density, $H_0$ is Hubble's constant, $\sigma_8$ is the amplitude of mass fluctuations, $\Omega_b$ is the baryon density as a fraction of the critical density, $n$~is the primordial spectral index, $\beta$ is the redshift distortion, $m_v$ is the neutrino mass, $\Gamma$~is \mbox{$\Omega_mH_0/100$ kms$^{-1}$Mpc$^{-1}$,} $\Omega_m^{0.6}\sigma_8$ is a combination of two other parameters that is useful in some peculiar velocity and lensing measurements, $\Omega_k$ is the curvature, and~$w_0$ is the equation of state for the dark energy parameter \citep{croft2015measurement}. For~this study, the~two parameters in question are $\sigma_8$ and $H_0$.

\subsection{Amplitude of Mass Fluctuations ($\sigma_8$)}

The amplitude of mass fluctuations ($\sigma_8$) is a parameter in the standard cosmological model that is concerned with the respective distributions of mass and light in the
Universe \citep{fan1997determining}. This is of interest to cosmologists because if $\sigma_8 \simeq 1$, the~implication is an "unbiased" Universe in which mass and light are evenly distributed in a sphere of radius $R$ = 8 $h^{-1}$ Mpc, whereas if $\sigma_8 \simeq 0.5$, the~result would be a "biased" Universe in which mass is distributed more extensively than light in a sphere of radius \mbox{$R $ = 8 h$^{-1}$} Mpc \citep{fan1997determining}. It is important for cosmologists to study and understand the distribution tendencies of mass and light in the Universe through $\sigma_8$ because large-scale differences in distribution of matter and energy in the present-day Universe tell us about density fluctuations in the early Universe on the cluster mass scale of $R$ = 8 h$^{-1}$ Mpc \citep{fan1997determining}.

\subsection{Hubble's Constant ($H_0$)}
Hubble's constant ($H_0$), like the amplitude of mass fluctuations, is a parameter in the standard cosmological~model.

\begin{figure}[H]
\centering\includegraphics[width=0.9\linewidth]{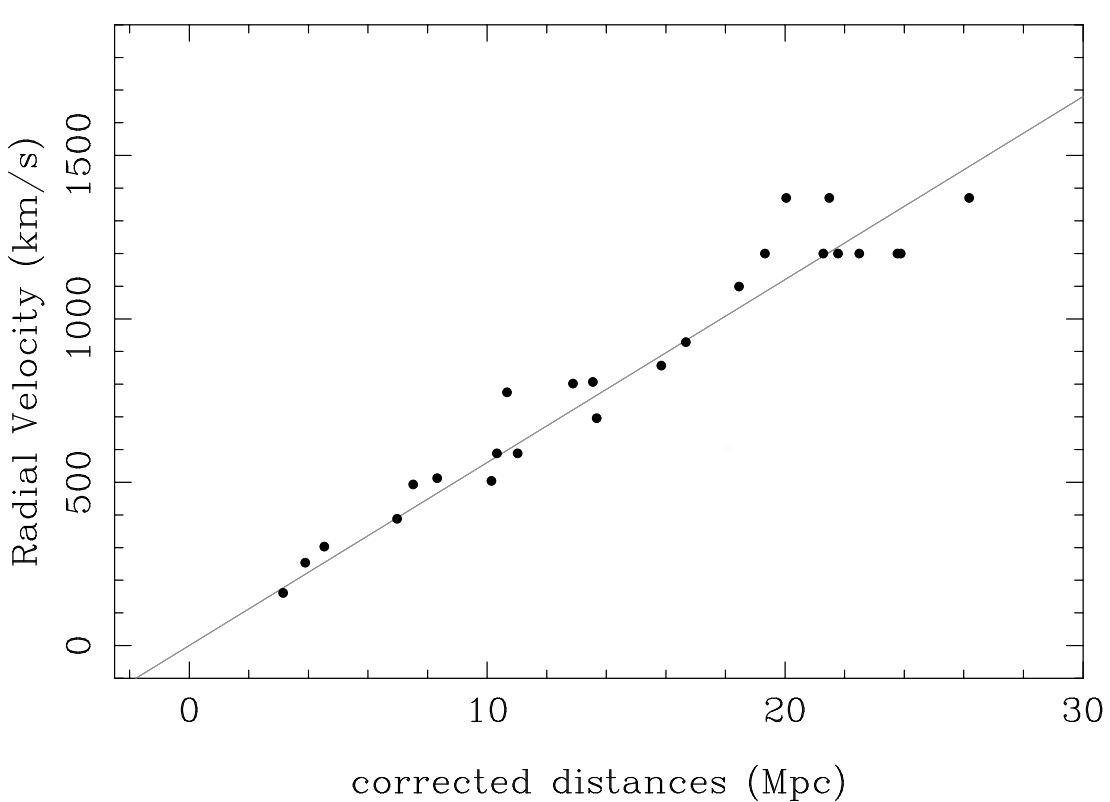}
\caption{{The Hubble--Lema\^itre Law \citep{elizalde2019reasons} representing radial recession velocity vs. distance from~observer.}}
\label{fig:1}
\end{figure}

$H_0$ is the slope of the line in the Hubble--Lema\^itre Law, relating the recession velocity of a galaxy to the distance that it is from an observer. A~representation of this law can be seen in Figure~\ref{fig:1}, obtained from \citet{georges2017hubble}. In~other words, $H_0$ relates to the expansion of the Universe on cosmic scales and is named after Edwin Hubble who discovered it in 1929 when he realized that galaxies' velocities away from an observer are directly proportional to their distance from that observer, except~for  cases of peculiar velocities \citep{kragh2003discovered}. In~recent years however, credit has also been given to Georges Lema\^itre jointly with Hubble for the discovery of this relationship \citep{elizalde2019reasons}. The~parameter is measured in km s$^{-1}$ Mpc$^{-1}$ and describes the velocity with which a galaxy of distance $d$ from an observer is moving radially away from that observer. 
Since the Universe is so large, these recession velocities in the form of redshift ($z$) are used to describe the distances to far away galaxies rather than units of length. Knowing the exact value of $H_0$ is important to cosmologists, as~$H_0$ can also be used to roughly calculate the age of the~Universe.

\subsection{Values and~Errors}

The first step in the process of determining the best observed values for the amplitude of mass fluctuations parameter ($\sigma_8$) and Hubble's constant ($H_0$) was to compile a list of several tens of measurements of these parameters. For~this specific project, 60 values were compiled for $\sigma_8$ between the years of 1993 and 2019 and 163 values were compiled for $H_0$ between the years of 1976 and 2019. In~addition to the values themselves, we were interested in a few other details about the measurements, namely, the years that those measurements were made in and the sizes of the error bars corresponding to the observed values. A~list of all 60 observed measurements for $\sigma_8$ 163 observed values for $H_0$ can be found in Tables \ref{tab:a1} and \ref{tab:a2}, respectively, in~the Appendix. For~$H_0$ values {(units throughout this paper in km s$^{1}$ Mpc$^{-1}$) between 1990 and 2010;  all of the values stem \mbox{from~\citet{croft2015measurement}}.} These~tables include the observed values along with their years of observation, sizes of error, and~references to  source articles. All of the referenced papers were found using the Astrophysics Data System (\url{https://ui.adsabs.harvard.edu/}), or~from the tables in \citet{croft2015measurement}. For~the statistical analysis of this data, a~simplifying assumption was made that each observed measurement is independent of the other observed measurements, eliminating the need for a covariance term. It should also be noted that the given error bars account for all statistical~effects.

\section{Statistical~Analysis}
\label{S:2}
\unskip
\subsection{Chi-Squared~Test}
\begin{figure}[H]
\centering\includegraphics[width=0.9\linewidth]{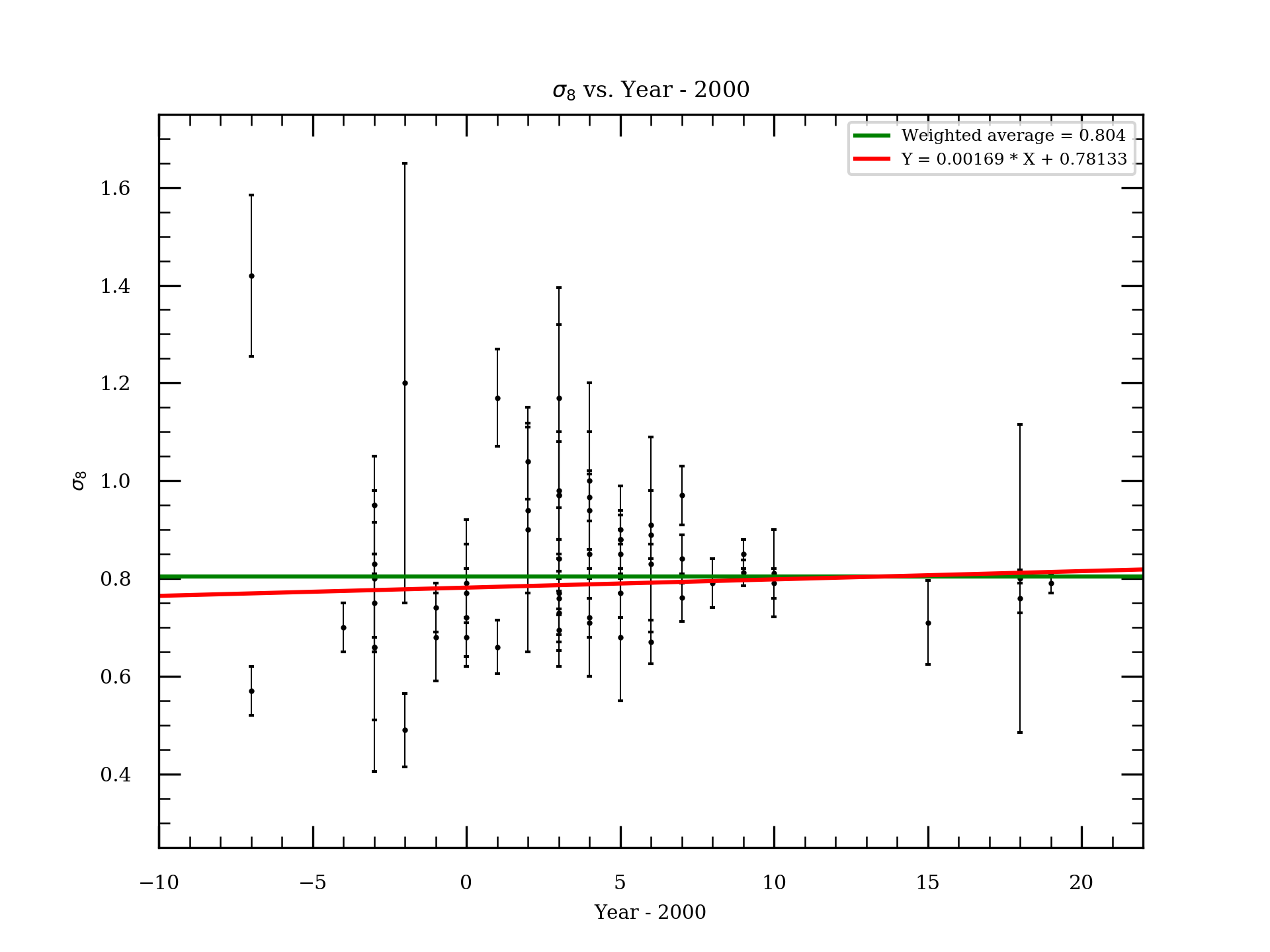}
\caption{{Data of $\sigma_8$ vs. time (year---2000) data, weighted average, and best linear fit.}}
\label{fig:2}
\end{figure}

\begin{figure}[H]
\centering\includegraphics[width=0.9\linewidth]{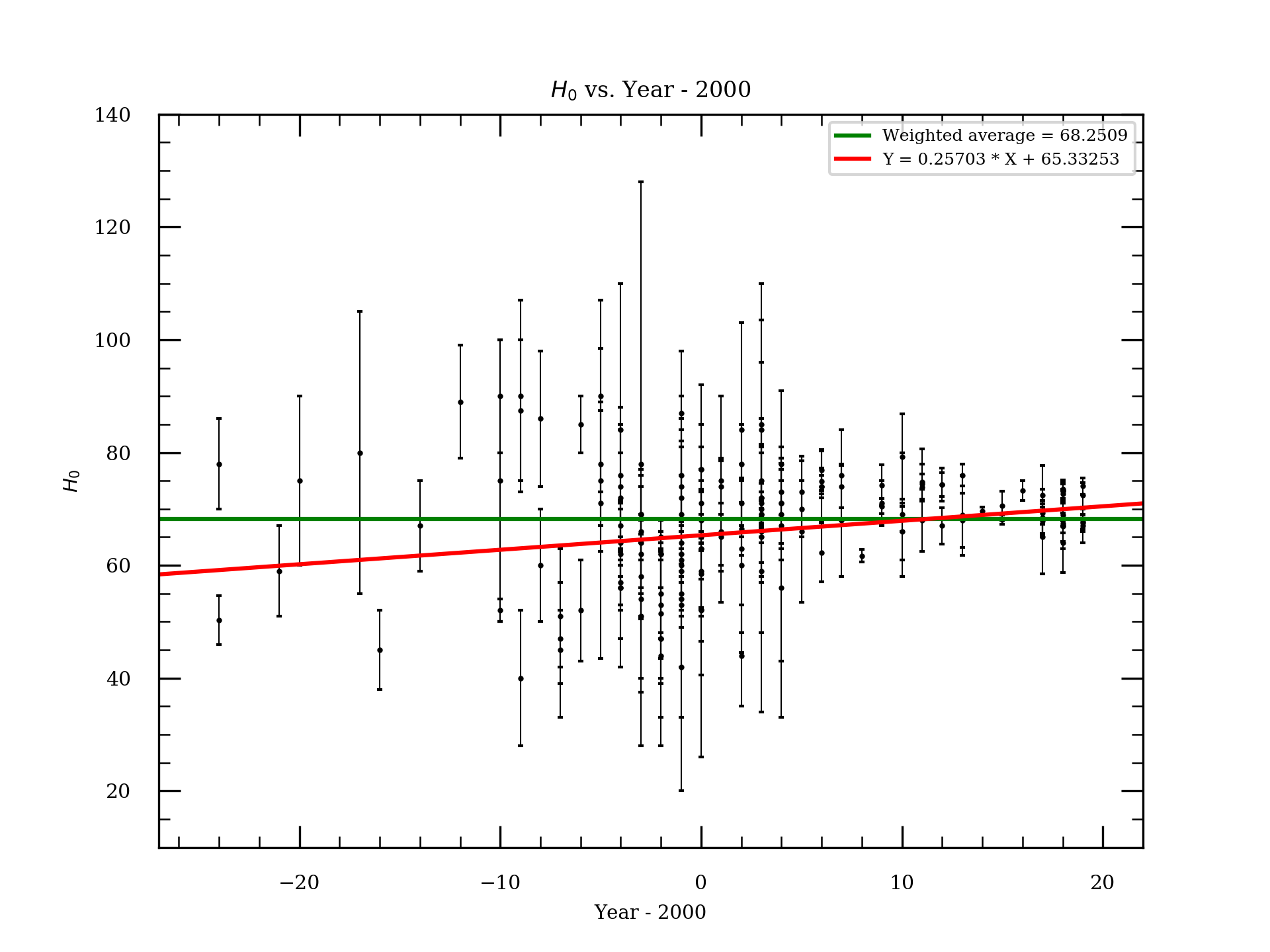}
\caption{{Data of $H_0$ vs. time (year---2000) data, weighted average, and best linear fit.}}
\label{fig:3}
\end{figure}

In order to analyze the trends in our datasets when viewed in scatter plots (see Figures~\ref{fig:2} and \ref{fig:3}), a~good statistical test is a chi-squared test. We used a chi-squared test to examine the probabilities of the deviations and determine whether the simplifying assumption made that the measurements were independent of one another was~correct.

\unskip

%\begin{figure}[h]
%\centering\includegraphics[width=0.9\linewidth]{sigma_8.png}
%\caption{$\sigma_8$ vs. Year - 2000}
%\end{figure}

%\begin{figure}[h]
%\centering\includegraphics[width=0.9\linewidth]{H_0.png}
%\caption{$H_0$ vs. Year - 2000. Units: km s$^{1}$ Mpc$^{-1}$.}
%\end{figure}

The chi-squared value of a set of data gives the likelihood that the trend observed in the data  occurred due to chance, and~is also known as a "goodness of fit" test \citep{plackett1983karl}. The~chi-squared value of a dataset is given by the following expression:
\begin{equation}
\label{Chisquared}
\chi^2 = \sum_{i=1}^{N}\frac{(x_{n,i} - x_{t,i})^2}{\sigma^2_i}
,\end{equation}
where in the case of our dataset $x_{n,i}$ is the observed value for the parameter, $x_{t,i}$ is the theoretical value for the parameter (weighted average or linear fit), $\sigma^2_i$ is the variance of the observed parameter value, {and $N$ is the number of points}. The~term for covariance term is absent from this expression due to the simplifying expression made that all of the observed measurements are independent of one another. This {independence} of data is precisely the hypothesis we want to test. {If the data were not independent, we would have to add a term for covariance to Equation~(\ref{Chisquared}). In any case, non-independency of our data would make the spread of the points lower than is indicated by the error bars, making the probability $Q$ (see Section~\ref{sec:2.3}) of higher deviations even lower, and~thus number of points to reject in order to have a distribution compatible to the error bars even larger. Therefore, our simplified approach can be considered a conservative calculation.}

This calculation was carried out twice, first using the weighted average $\sigma_8$ and $H_0$ values as the theoretical values ($x_t$), and~then again using the best fit values from a linear fit designed to minimize the value of $\chi^2$ as $x_t$. Lines representing both the weighted average of the dataset (blue) and the best fit for the dataset (red) that were used to calculate chi-squared can be seen with the data points {in \mbox{Figures \ref{fig:2} and \ref{fig:3}}}. The~weighted averages ($\lambda_w$) of the parameters in question were calculated by weighting each point by the variance of that value, as~shown below, where $\sigma^2_i$ is the variance of data point $i$:
\begin{equation}
\label{Weighted_avg}
   \lambda_w = \frac{\sum_{i=1}^{n}\frac{x_i}{\sigma^2_i}}{\sum_{i=1}^{n}\frac{1}{\sigma^2_i}}
\end{equation}

For $\sigma_8$, $\lambda_w \approx 0.8038$ and for $H_0$, $\lambda_w \approx 69.3815$. Substituting these weighted averages in for $x_t$ in Equation~(\ref{Chisquared}) gives $\chi^2 \approx 189.037$ for $\sigma_8$ and $\chi^2 \approx 575.655$ for $H_0$.

In order to find the linear fit of the form:
\begin{equation}
\label{y:ax+b}
   Y = A + B \times X
\end{equation}
%\begin{flushleft}
where $Y$ is the theoretical value for the parameter being analyzed and $X$ is the year of that measurement minus 2000. A~program was written in Python that minimizes $\chi^2$. {When replacing $Y$ from Equation~(\ref{sigma_fit}) for $x_t$ in Equation~(\ref{Chisquared}), we found that $\chi^2 \approx 182.4$ for $\sigma_8$ and $\chi^2 \approx 480.1$ for $H_0$.}
\mbox{In order to} calculate the error bars for the parameters \emph{A} and \emph{B}, a~program was written in Python to estimate the range of values for $\sigma_8$ and $H_0$ with an error of 1$\sigma$ added. {The 1$\sigma$ error (68\% C.L.) was obtained by adding the value of $2.3\left(\frac{\chi ^2}{n}\right)$ to the minimum of $\chi^2$ values of 182.4 ($\sigma_8$) and 480.1 ($H_0$) in accordance to the process followed in \citet{avni1976energy},
where $n$ is the number of degrees of freedom 
and the second factor was added to account for either under or overestimation of the error bars. For~our $\sigma _8$ values, this process resulted in an A value of $0.781\pm 0.012$ and a $B$ value of $(1.7\pm 0.8) \times 10^{-3}$.} With these values for $A$ and $B$, the~function of the linear fit for $\sigma_8$ becomes:
%\end{flushleft}
{\bf
\begin{equation}
\label{sigma_fit}
   Y = 0.781 + (1.7 \times 10^{-3}) \times X
\end{equation}
}

%\newpage

{For the $H_0$ values, this process resulted in an $A$ value of $65.3\pm 0.6$ and a $B$ value of $0.26\pm 0.04$, making the function of the linear fit for $H_0$:
\begin{equation}
\label{hubble_fit}
   Y = 65.3 + 0.26 \times X
\end{equation}
}
\begin{flushleft}
as can be seen {in Figures~\ref{fig:2} and \ref{fig:3}}, represented by the red line.
\end{flushleft}

\subsection{Reduced~Chi-Squared}

In order to account for the degrees of freedom in the data, a~reduced chi-squared test was used to test the goodness of fit for both the weighted average and best fit values. Reduced chi-squared is commonly used for several purposes in astronomy, namely, model comparison and error estimation~\citep{andrae2010and}. The~reduced chi-squared value of a dataset is simply the chi-squared value divided by the degrees of freedom ($n$) of that dataset, as~shown in the following relation:
\begin{equation}
\label{Redchisq}
   \chi^2_n = \frac{\chi^2}{n}
\end{equation}

In the case of this analysis, for~the weighted average calculations there were 59 degrees of freedom for $\sigma_8$ and 162 degrees of freedom for $H_0$ (one free parameter). For~the linear fit calculations there were 58 degrees of freedom for $\sigma_8$ and 161 degrees of freedom for $H_0$ (two free parameters). When applying the $\chi^2$ value calculated using the weighted average of the dataset to Equation~(\ref{hubble_fit}), we get a reduced chi-squared (or, chi-squared per degree of freedom) of 3.20 for $\sigma_8$ and a reduced chi-squared value of 3.55 for $H_0$. {Likewise, the~reduced chi-squared value obtained from the best fit function meant to minimize reduced chi-squared is 3.04 for $\sigma_8$ and is 2.95 for $H_0$, both of which, in~accordance to theory, are less than those calculated using the weighted average (0.16 difference for $\sigma_8$ and 0.60 for $H_0$).}

\subsection{Statistical Significance, Q} 
\label{sec:2.3}
The probability that a calculated $\chi^2$ value for a dataset with $n$ degrees of freedom is due to chance is represented by $Q$ and is given by the following expression:
\begin{equation}
   Q_{\chi^2,n} = [2^{d/2}\Gamma(\frac{d}{2})]^{-1}\int_{\chi^2}^{\infty}(t)^{\frac{d}{2}-1}e^{\frac{-t}{2}}dt
\end{equation}

\begin{flushleft}
where $\Gamma_x$ is given by:
\end{flushleft}
\begin{equation}
   \Gamma_x = \int_{0}^{\infty}t^{x-1}e^{-t}dt
\end{equation}
%\begin{flushleft}
and is known as the generalization of the factorial function to real and complex arguments \citep{gronau2003gamma}. \mbox{In order to} determine which values should be removed as bad values, all values were ranked based on their contributions to $\chi^2$ by increasing value of [$x -$ (best fit $x$)]/(error of $x$) and then again by \mbox{[$x -$ (weighted average $x$)]/(error of $x$),} where x is the observed value for the parameter in question. Values with the largest contribution to $\chi^2$ (bad values) were removed first.
%\end{flushleft}

\subsubsection{Amplitude of Mass~Fluctuations}

\begin{figure}[H]
\centering\includegraphics[width=0.9\linewidth]{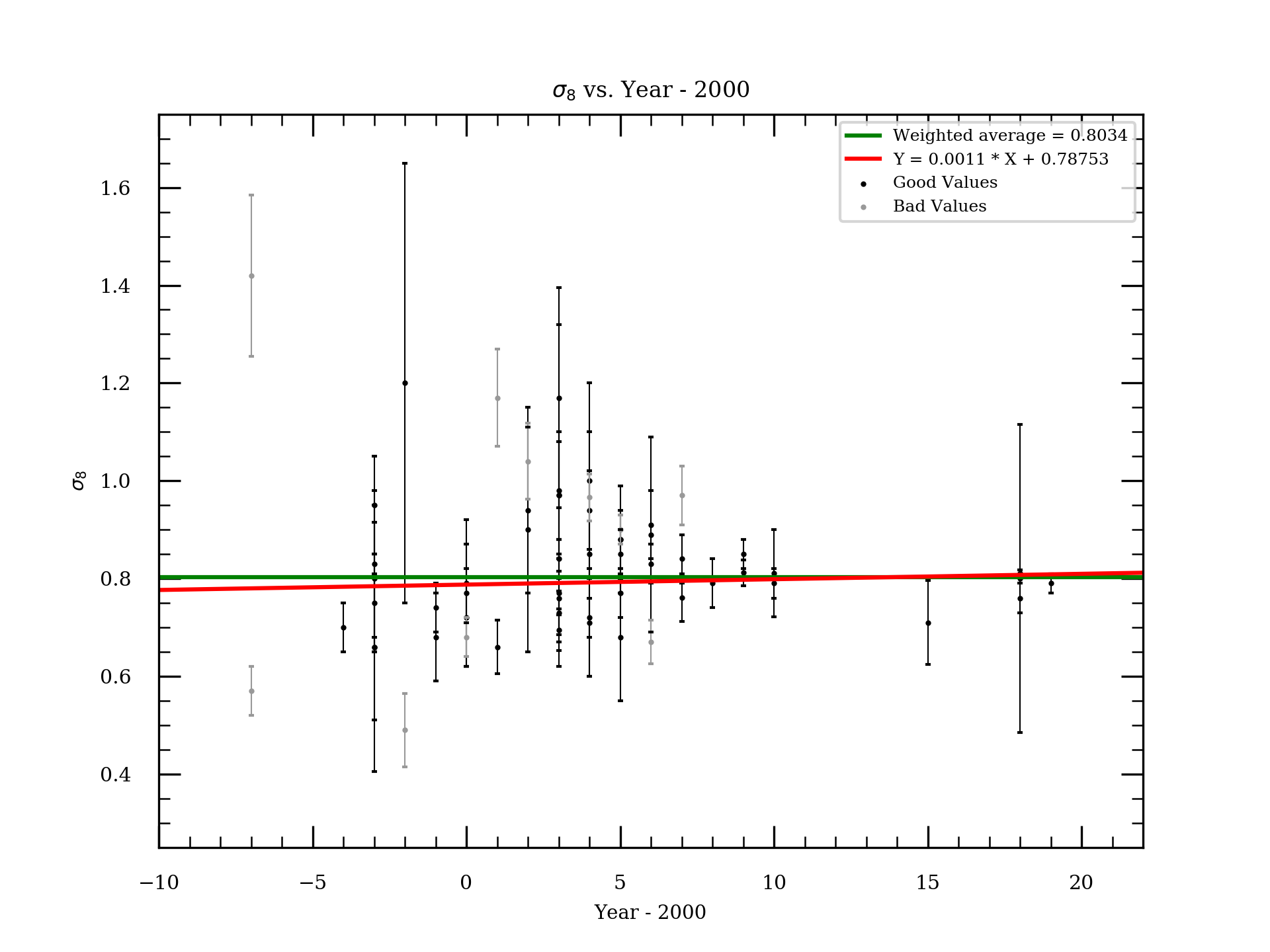}
\caption{{Data of $\sigma _8$ vs. time (year---2000) separating the $N=50$ good 
values that make the $\chi ^2$ linear fit compatible with the error bars, and~the
 rest of the points ($N=10$) plotted as bad values. Here, we only used the 
good values for the weighted average and best linear fit.}}
\label{fig:4}
\end{figure}

For the value of $\chi^2$ calculated using the weighted average of $\sigma_8$ ($n = 59, \chi^2 \approx 189.0$), the~probability that the observed trend is due to chance is $Q = 1.6 \times 10^{-15}$. In~order to reach a value for $Q$ that is statistically significant ($Q \geq 0.05$), 14 bad values must be removed from the data ($n = 45, \chi^2 \approx 58.1548$), producing a value for $Q$ of 0.0902. {For the value of $\chi^2$ calculated using the best fit function designed to minimize $\chi^2$ ($n = 58, \chi^2 \approx 182.4$), $Q = 8.8 \times 10^{-15}$. In~order to reach a statistically significant value
for $Q$, 10 bad values must be removed from the data ($n = 48, \chi^2 \approx 61.0$), producing a value for $Q$ of 0.099. With~this last subsample of 50 points, the~best linear fit of $\sigma _8$ returned an $A$ value of $0.787\pm 0.008$ and a $B$ value of $(1.1\pm 0.5)\times 10^{-3}$; see Figure~\ref{fig:4}.}

%\unskip

\subsubsection{Hubble's~Constant}
For the value of $\chi^2$ calculated using the weighted average of $H_0$ ($n = 162, \chi^2 \approx 575.655$), the~probability that the observed trend is due to chance is $Q = 1.0 \times 10^{-47}$. In~order to reach a value for $Q$ that is statistically significant ($Q \geq 0.05$), 36 bad values must be removed from the data ($n = 125, \chi^2 \approx 152.5541$), producing a value for $Q$ of 0.0538. {For the value of $\chi^2$ ($n = 161, \chi^2 \approx 480.1$) calculated using the best fit function designed to minimize $\chi^2$, $Q = 1.8 \times 10^{-33}$. In~order to reach a statistically significant value for $Q$, 24 bad values must be removed ($n = 137, \chi^2 \approx 164.1$), producing a value for $Q$ of 0.057.
With this last subsample of 139 points, the~best linear fit of $H_0$ returned an $A$ value of $65.9\pm 0.4$ and a $B$ value of $0.277^{+0.032}_{-0.034}$; see Figure~\ref{fig:5}.

\begin{figure}[H]
\centering\includegraphics[width=0.9\linewidth]{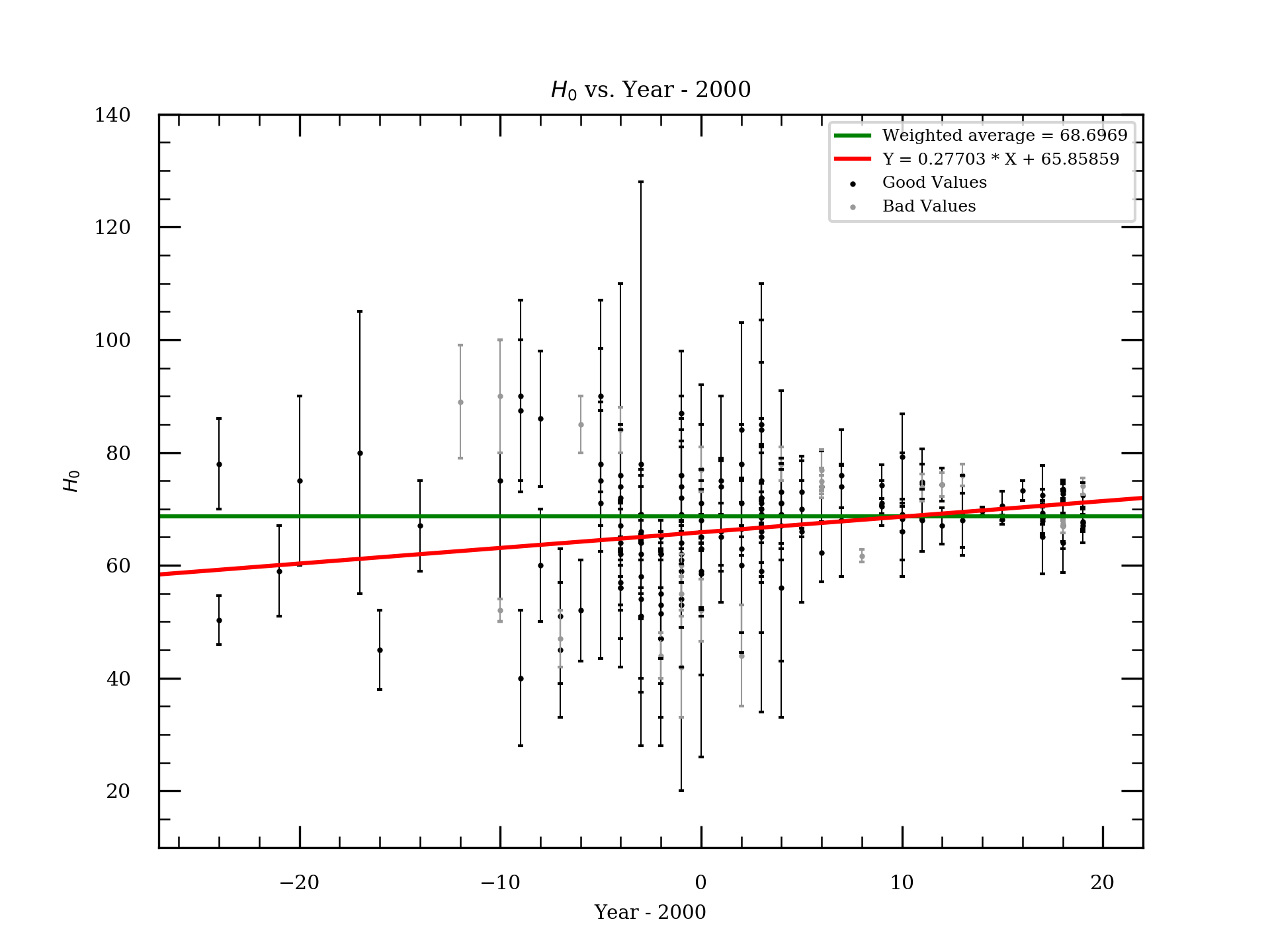}
\caption{{Data of $H_0$ vs. time (year---2000) separating the $N=139$ good values that make the $\chi ^2$ linear fit compatible with the error bars, and~the
 rest of the points ($N=24$) plotted as bad values. Here,~we~only use the good values for the weighted average and best linear fit.}}
\label{fig:5}
\end{figure}
%\unskip

The non-zero value of $B$ is very significant; however, the error of $B$ may be 
non-Gaussian and we cannot directly interpret this as significant evolution. The~correlation factor of $H_0$ with time\footnote{{For two independent variables $X$ and $Y$,  the~correlation factor 
is defined as $c=\frac{\langle X\, Y\rangle}{\langle X\rangle \langle Y\rangle}-1$, with~error $Err(c)=\frac{\sigma _X\sigma _Y}{\sqrt{N}\langle X\rangle \langle Y\rangle}$. 
The Pearson correlation coefficient would be $\frac{c}{\sqrt{N}Err(c)}$.}}  
is $c=0.027\pm 0.013$, a~$2\sigma $ significant correlation.}

\section{Conclusions and~Discussion}

The original $Q$ values for both the weighted average and best fit calculations of the probability of the data for both parameters are extremely low before the removal of bad values. Even though this is the case, a~rather large discrepancy can be seen in how many bad values need removing to reach a statistically significant dataset ($Q\geq 0.05$). 
{For the $\sigma_8$ values, to~attain statistical significance, the~weighted average calculation needs 14 bad values removed, whereas the best fit calculation needs only 10 bad values removed. For~the $H_0$ values, to~attain statistical significance, the~weighted average calculation requires 36 bad values be removed, whereas the best fit calculation only needs 24 bad values removed.} With the studies of both parameters ending in the aforementioned conclusions, it~is reasonable to conclude that the linear fit with time (year---2000) on the $x-$axis and measurements of the parameters in question ($\sigma_8$ and $H_0$) on the $y-$axis is a better estimation of the data than the weighted averaged of the data weighted with the inverse square proportion of the error of each value in question, a~linear fit is a better estimate of the data than the weighted~average.

{For $H_0$, we observed a slight growing trend (at 2-$\sigma $ level) in the value of the measurements in the last 43 years, although~the interpretation of this upward trend as a random fluctuation is not excluded.}
%This statistical analysis of the historical evolution of cosmological parameters {\bf reinforces} the words by \citet{feynman1974cargo}, where Richard Feynman gives an example of Nobel Laureate Robert Millikan measuring the charge of an electron: "it's interesting to look at the history of measurements of the charge of the
%electron, after Millikan. If you plot them as a function of time, you find that one is a little
%bigger than Millikan's, and the next one is a little bit bigger than that, and the next ones a
%little bit bigger than that, until finally they settle down to a number which is higher"\citep{feynman1974cargo}. Feynman goes on to ask why the final higher number was not discovered right away, and comes to the conclusion that when "[scientists] got a number that was too high above Millikan’s, they thought something must be wrong
%and they would look for and find a reason why something might be wrong, leading them to eliminate values that were too far off, and did other things like that". We have seen here that
%the case of the charge of the electron happens also within cosmology.
%Improved research and observations have led to greater precision in measurements of parameters in the standard cosmological model, and research in this field should not only be continued, but encouraged, in order
%to achieve the epoch of stable values of the parameter independent of fashions.

In addition to the increasing precision of measurements, it is concluded from this analysis that the error bars of the observed parameters have been largely underestimated in at least 20\% of the measurements, or~the systematic errors of the observation techniques were not fully considered. It~should also be stated that, due to the simplifying assumption about the covariance of each observed measurement, 20\% of the error bars being underestimated is a conservative percentage (in reality, it is a minimum of 20\% the measurements). In~the light of the analysis carried out in this paper, one would not be {surprised} to find cases like
the 4.4$\sigma$ discrepancy seen between the best measurement using Supernovae Ia in \citet{riess2019large} of $H_0$ = $74.03\pm1.42$ km s$^{-1}$ Mpc$^{-1}$ and the value derived from cosmic microwave background radiation of $H_0$ = $67.4\pm0.5$ km s$^{-1}$ Mpc$^{-1}$. It is likely that the underestimation of error bars for $H_0$ in many measurements contributes to the apparent 4.4$\sigma$ discrepancy formally known as the Hubble~tension.

\vspace{6pt}
\authorcontributions{Conceptualization, T.F. and M.L.-C.; methodology, M.L.-C.; software, T.F.; formal analysis, T.F.; writing---original draft preparation, T.F.; writing---review and editing, M.L.-C.; supervision, M.L.-C. All authors have read and agreed to the published version of the~manuscript.}

\funding{MLC was supported by the grant PGC-2018-102249-B-100 of the Spanish 
Ministry of Economy and Competitiveness (MINECO).} %MDPI: Please add: "This research received no external funding" or "This research was funded by NAME OF FUNDER grant number XXX." and  and "The APC was funded by XXX". Check carefully that the details given are accurate and use the standard spelling of funding agency names at \url{https://search.crossref.org/funding}, any errors may affect your future funding.

\acknowledgments{Thanks are given to Martin Sahlen and Andreas Korn for their suggestions for this work, and~to Rupert
Croft for providing data of his paper  \citet{croft2015measurement}. {Thanks are given to the two anonymous
referees for helpful comments.}
}

\conflictsofinterest{The authors declare no conflict of~interest.}

%\bibliographystyle{mdpi} 
%\bibliography{bib}
%
%\onecolumn

\appendix
\appendixtitles{yes}
\section{Tables of~Data}

\begin{table}[H]
\caption{$\sigma_8$ data.}
\centering
\scalebox{.95}[0.95]{ \begin{tabular}{cccc}
  \toprule
 \textbf{Date} & \boldmath{$\sigma_8$} & \boldmath{$\pm$} & \textbf{Reference} \\ [0.5ex]
 \midrule
 1993 & 0.57 & 0.05 &  \citet{white1993amplitude} \\
 \midrule
 1993 & 1.415 & 0.165 & \citet{white1993amplitude} \\
 \midrule
 1996 & 0.7 & 0.05 & \citet{taylor1996non} \\
 \midrule
 1997 & 0.75 & 0.1 &  \citet{carlberg1997cnoc} \\
 \midrule
 1997 & 0.95 & 0.1 &   \citet{carlberg1997cnoc} \\
 \midrule
 1997 & 0.8 & 0.15 &   \citet{shimasaku1997measuring} \\
 \midrule
 1997 & 0.66 & $+0.22$ $-0.08$ &  \citet{henry1997measurement} \\
 \midrule
 1997 & 0.66 & $+0.34$ $-0.17$ & \citet{henry1997measurement} \\
 \midrule
 1997 & 0.83 & 0.15 &  \citet{fan1997determining} \\
 \midrule
 1998 & 1.2 & $+0.5$ $-0.4$ &  \citet{bahcall1998most}  \\
 \midrule
 1998 & 0.49 & $+0.08$ $-0.07$ &  \citet{robinson1998simultaneous} \\
 \midrule
 1999 & 0.68 & 0.09 &  \citet{einasto1999steps} \\
 \midrule
 1999 & 0.74 & 0.05 &  \citet{bridle1999cosmological} \\
 \midrule
 2000 & 0.72 & 0.1 &  \citet{henry2000measuring} \\
 \midrule
 2000 & 0.77 & 0.15 & \citet{henry2000measuring} \\
 \midrule
 2000 & 0.79 & 0.08 &  \citet{matsubara2000cosmological} \\
 \midrule
 2000 & 0.68 & 0.04 &  \citet{mcdonald2000observed} \\
 \bottomrule
2001 & 1.17 & 0.1 &  \citet{bridle2000cosmological} \\
 \midrule
 2001 & 0.66 & $+0.06$ $-0.05$ &  \citet{borgani2001measuring} \\
 \midrule
 2002 & 0.94 & 0.17 &  \citet{refregier2002cosmic} \\
 \midrule
 2002 & 1.04 & 0.104 &  \citet{evrard2002galaxy} \\
 \midrule
 \end{tabular} }
 \end{table}

\begin{table}[H]\ContinuedFloat
\centering
\caption{{\em Cont.}}
\scalebox{.95}[0.95]{\begin{tabular}{cccc}
\toprule
 \textbf{Date} & \boldmath{$\sigma_8$} & \boldmath{$\pm$} & \textbf{Reference} \\ [0.5ex]
 \midrule
 
 2002 & 1.04 & 0.078 &  \citet{komatsu2002sunyaev} \\
 \midrule
 2002 & 0.9 & $+0.3$ $-0.2$ &  \citet{bahcall2002cluster} \\
 \midrule
 2003 & 0.76 & 0.09 &  \citet{melchiorri2003cosmological} \\
 \midrule
 2003 & 0.98 & 0.1 &  \citet{bahcall2003amplitude} \\
 \midrule
 2003 & 0.73 & $+0.06$ $-0.03$ &  \citet{brown2003shear} \\
 \midrule
 2003 & 1.17 & $+0.25$ $-0.2$ &  \citet{slosar2003cosmological} \\
 \midrule
 2003 & 0.77 & $+0.05$ $-0.04$ &  \citet{pierpaoli2003determining} \\
 \midrule
 2003 & 0.695 & 0.042 &  \citet{allen2003cosmological} \\
 \midrule
 2003 & 0.84 & 0.04 &  \citet{spergel2003first} \\
 \midrule
 2003 & 0.97 & 0.13 &  \citet{bacon2003joint} \\
 \midrule
 2003 & 0.97 & 0.35 &  \citet{hamana2003cosmic} \\
 \midrule
 2004 & 0.966 & 0.048 &  \citet{pope2004cosmological} \\
  \midrule
 2004 & 0.71 & 0.11 &  \citet{heymans2004weak} \\
 \midrule
 2004 & 0.72 & 0.04 &  \citet{voevodkin2004constraining} \\
 \midrule
 2004 & 0.85 & $+0.38$ $-0.12$ &  \citet{lokas2004cluster} \\
 \midrule
 2004 & 0.94 & 0.08 &  \citet{lokas2004cluster} \\
 \midrule
 2004 & 1.0 & 0.2 &  \citet{chang2004weak} \\
 \midrule
 2005 & 0.90 & 0.03 &  \citet{seljak2005cosmological} \\
 \midrule
 2005 & 0.88 & 0.06 &  \citet{seljak2005sdss} \\
 \midrule
 2005 & 0.68 & 0.13 &  \citet{heymans2005cosmological} \\
 \midrule
 2005 & 0.85 & 0.05 &  \citet{pike2005cosmological} \\
 \midrule
 2005 & 0.88 & $+0.12$ $-0.10$ &  \citet{gaztanaga2005statistical} \\
 \midrule
 2006 & 0.89 & 0.2 &  \citet{eke2006galaxy} \\
 \midrule
 2006 & 0.77 & 0.05 &  \citet{sanchez2006cosmological} \\
 \midrule
 2006 & 0.91 & 0.07 &  \citet{viel2006cosmological} \\
 \midrule
 2006 & 0.67 & $+0.04$ $-0.05$ &  \citet{dahle2006cluster} \\
 \midrule
 2007 & 0.761 & $+0.049$ $-0.048$ &  \citet{spergel2007three} \\
 \midrule
 2007 & 0.84 & 0.05 &  \citet{benjamin2007cosmological} \\
 \midrule
 2007 & 0.97 & 0.06 &  \citet{harker2007constraints} \\
 \midrule
 2008 & 0.79 & 0.05 &  \citet{ross2008normalization} \\
 \midrule
 2009 & 0.85 & $+0.04$ $-0.02$ & \citet{henry2009x} \\
 \midrule
 2009 & 0.812 & 0.026 & \citet{komatsu2009five} \\
 \midrule
 2010 & 0.79 & 0.03 &  \citet{mantz2010observed} \\
 \midrule
 2010 & 0.811 & 0.089 &  \citet{hilbert2010abundances} \\
 \midrule
 2014 & 0.83 & 0.04 &  \citet{mantz2014weighing} \\
 \midrule
 2015 & 0.710 & 0.086 &  \citet{gil2015power} \\
 \midrule
 2018 & 0.811 & 0.006 &  \citet{aghanim2018planck} \\
 \midrule
 2018 & 0.76 & 0.03 &  \citet{salvati2018constraints} \\
 \midrule
 2018 & 0.80 & 0.31 &  \citet{corasaniti2018probing} \\
 \midrule
 2019 & 0.786 & 0.02 &  \citet{kreisch2019neutrino} \\
 \bottomrule
\end{tabular}}
\label{tab:a1}
\end{table}

%\eject

\begin{table}[H]
\caption{$\sigma_8$ data.}
\centering
 \begin{tabular}{cccc}
  \toprule
   \textbf{Date} & \boldmath{$H_0$} \textbf{(km s}\boldmath{$^{-1}$} \textbf{Mpc}\boldmath{$^{-1}$}\textbf{)} & \boldmath{$\pm$} & \textbf{Reference} \\ [0.5ex]
 \midrule
 1976 & 78 & 8 &  \citet{jaakkola1976remarks} \\
 \midrule
 1976 & 50.3 & 4.3 &  \citet{sandage1976steps} \\
 \midrule
 1979 & 59 & 8 &  \citet{visvanathan1979distance} \\
 \midrule
 1980 & 75 & 15 &  \citet{stenning1980local} \\
 \midrule
 1983 & 80 & 25 & \citet{rubin1983new} \\
 \midrule
 1984 & 45 & 7 &  \citet{joeveer1984type} \\
 \midrule
 1986 & 67 & 8 & \citet{gondhalekar1986ultraviolet} \\
 \midrule
 1988 & 89 & 10 & \citet{melnick1988giant} \\
 \midrule
 1990	&	90	&	10	&  \citet{croft2015measurement}	\\
 \midrule
1990	&	75	&	25	&	 \citet{croft2015measurement}	\\
\midrule
1990	&	52	&	2	&	 \citet{croft2015measurement}	\\
\midrule
1991	&	90	&	17	&	 \citet{croft2015measurement}	\\
\midrule
1991	&	87.5	&	12.5	&	 \citet{croft2015measurement}	\\
\midrule
1991	&	40	&	12	&	 \citet{croft2015measurement}	\\
\midrule
1992	&	86	&	12	&	 \citet{croft2015measurement}	\\
\midrule
1992	&	60	&	10	&	 \citet{croft2015measurement}	\\
\midrule
1993	&	51	&	12	&	 \citet{croft2015measurement}	\\
\midrule
1993	&	47	&	5	&	 \citet{croft2015measurement}	\\
\midrule
1993	&	45	&	12	&	 \citet{croft2015measurement}	\\
\midrule
1994	&	85	&	5	&	 \citet{croft2015measurement}	\\
\midrule
1994	&	52	&	9	&	 \citet{croft2015measurement}	\\
\midrule
1995	&	93	&	1	&	 \citet{croft2015measurement}	\\
\midrule
1995	&	90	&	17	&	 \citet{croft2015measurement}	\\
\midrule
1995	&	78	&	11	&	 \citet{croft2015measurement}	\\
\midrule
1995	&	75	&	12.5	&	 \citet{croft2015measurement}	\\
\midrule
1995	&	71	&	27.5	&	 \citet{croft2015measurement}	\\
\midrule
1996	&	84	&	4	&	 \citet{croft2015measurement}	\\
\midrule
1996	&	76	&	34	&	 \citet{croft2015measurement}	\\
\midrule
1996	&	74	&	11	&	 \citet{croft2015measurement}	\\
\midrule
1996	&	72	&	12	&	 \citet{croft2015measurement}	\\
\midrule
1996	&	67	&	4.5	&	 \citet{croft2015measurement}	\\
\midrule
1996	&	64	&	6	&	 \citet{croft2015measurement}	\\
\midrule
1996	&	62	&	9	&	 \citet{croft2015measurement}	\\
\midrule
1996	&	57	&	4	&	 \citet{croft2015measurement}	\\
\midrule
1996	&	56	&	4	&	 \citet{croft2015measurement}	\\
\midrule
 1996	&	56	&	9	&	 \citet{croft2015measurement}	\\
\midrule
1997	&	78	&	50	&	 \citet{croft2015measurement}	\\
\midrule
1997	&	69	&	5	&	 \citet{croft2015measurement}	\\
\midrule
1997	&	69	&	8	&	 \citet{croft2015measurement}	\\
\midrule
1997	&	66	&	10	&	 \citet{croft2015measurement}	\\
\midrule
1997	&	64	&	13	&	 \citet{croft2015measurement}	\\
 \bottomrule
 \end{tabular}
 \end{table}

\begin{table}[H]\ContinuedFloat
\centering
\caption{{\em Cont.}}
\begin{tabular}{cccc}
\toprule
 \textbf{Date} & \boldmath{$H_0$} \textbf{(km s}\boldmath{$^{-1}$} \textbf{Mpc}\boldmath{$^{-1}$}\textbf{)} & \boldmath{$\pm$} & \textbf{Reference} \\ [0.5ex]
\midrule
1997	&	62	&	7	&	 \citet{croft2015measurement}	\\
\midrule
1997	&	58	&	7.5	&	 \citet{croft2015measurement}	\\
\midrule
1997	&	54	&	14	&	 \citet{croft2015measurement}	\\
\midrule
1997	&	51	&	13.5	&	 \citet{croft2015measurement}
\\
 \midrule
1998	&	65	&	1	&	 \citet{croft2015measurement}	\\
\midrule
1998	&	62	&	6	&	 \citet{croft2015measurement}	\\
\midrule
1998	&	62	&	6	&	 \citet{croft2015measurement}	\\
\midrule
1998	&	55	&	8	&	 \citet{croft2015measurement}	\\
\midrule
1998	&	53	&	9.5	&	 \citet{croft2015measurement}	\\
\midrule
1998	&	51.5	&	12.5	&	 \citet{croft2015measurement}	\\
\midrule
1998	&	47	&	19	&	 \citet{croft2015measurement}	\\
\midrule
1998	&	47	&	14	&	 \citet{croft2015measurement}	\\
\midrule
1998	&	44	&	4	&	 \citet{croft2015measurement}	\\
\midrule
1999	&	87	&	11	&	 \citet{croft2015measurement}	\\
\midrule
1999	&	76	&	14	&	 \citet{croft2015measurement}	\\
\midrule
1999	&	74	&	8	&	 \citet{croft2015measurement}	\\
\midrule
1999	&	72	&	9	&	 \citet{croft2015measurement}	\\
\midrule
1999	&	69	&	15	&	 \citet{croft2015measurement}	\\
\midrule
1999	&	64	&	3.75	&	 \citet{croft2015measurement}
\\
\midrule
1999	&	62	&	5	&	 \citet{croft2015measurement}	\\
\midrule
1999	&	61	&	7	&	 \citet{croft2015measurement}	\\
\midrule
1999	&	60	&	2	&	 \citet{croft2015measurement}	\\
\midrule
1999	&	59	&	17	&	 \citet{croft2015measurement}	\\
\midrule
1999	&	55	&	3	&	 \citet{croft2015measurement}	\\
\midrule
1999	&	54	&	5	&	 \citet{croft2015measurement}	\\
\midrule
1999	&	53	&	33	&	 \citet{croft2015measurement}	\\
\midrule
1999	&	42	&	9	&	 \citet{croft2015measurement}	\\
\midrule
2000	&	77	&	8	&	 \citet{croft2015measurement}	\\
\midrule
2000	&	77	&	4	&	 \citet{croft2015measurement}	\\
\midrule
 2000	&	71	&	6	&	 \citet{croft2015measurement}	\\
\midrule
2000	&	68	&	5.4	&	 \citet{croft2015measurement}	\\
\midrule
2000	&	65	&	1	&	 \citet{croft2015measurement}	\\
\midrule
2000	&	63	&	10.5	&	 \citet{croft2015measurement}
\\
\midrule
2000	&	63	&	12	&	 \citet{croft2015measurement}	\\
\midrule
2000	&	59	&	33	&	 \citet{croft2015measurement}	\\
\midrule
2000	&	58.5	&	6.3	&	 \citet{croft2015measurement}
\\
\midrule
2000	&	52.2	&	11.65	&	 \citet{croft2015measurement}
\\
\midrule
2000	&	52	&	5.5	&	 \citet{croft2015measurement}	\\
\midrule
2001	&	75	&	15	&	  \citet{croft2015measurement}	\\
\midrule
2001	&	74	&	5	&	 \citet{croft2015measurement}	\\
 \bottomrule
 \end{tabular}
 \end{table}

\begin{table}[H]\ContinuedFloat
\centering
\caption{{\em Cont.}}
\begin{tabular}{cccc}
\toprule
 \textbf{Date} & \boldmath{$H_0$} \textbf{(km s}\boldmath{$^{-1}$} \textbf{Mpc}\boldmath{$^{-1}$}\textbf{)} & \boldmath{$\pm$} & \textbf{Reference} \\ [0.5ex]
\midrule
2001	&	66	&	12.5	&	 \citet{croft2015measurement}
\\
\midrule
2001	&	65	&	6	&	 \citet{croft2015measurement}	\\ \midrule
2002	&	84	&	19	&	 \citet{croft2015measurement}	\\
\midrule
2002	&	78	&	7	&	 \citet{croft2015measurement}	\\
\midrule
2002	&	71	&	4	&	 \citet{croft2015measurement}	\\
\midrule
2002	&	66.5	&	4.7	&	 \citet{croft2015measurement}
\\
\midrule
2002	&	63	&	15	&	 \citet{croft2015measurement}	\\
\midrule
2002	&	60	&	15.5	&	 \citet{croft2015measurement}
\\
\midrule
2002	&	44	&	9	&	 \citet{croft2015measurement}	\\
 \midrule
2003	&	85	&	18.5	&	 \citet{croft2015measurement}
\\
\midrule
2003	&	84	&	26	&	 \citet{croft2015measurement}	\\
\midrule
2003	&	75	&	6.5	&	 \citet{croft2015measurement}	\\
\midrule
2003	&	72	&	14	&	 \citet{croft2015measurement}	\\
\midrule
2003	&	72	&	8	&	 \citet{croft2015measurement}	\\
\midrule
2003	&	71	&	3.5	&	 \citet{croft2015measurement}	\\
\midrule
2003	&	70	&	3	&	 \citet{croft2015measurement}	\\
\midrule
2003	&	69	&	12	&	 \citet{croft2015measurement}	\\
\midrule
2003	&	69	&	4	&	 \citet{croft2015measurement}	\\
\midrule
2003	&	68.4	&	1.7	&	 \citet{croft2015measurement}
\\
\midrule
2003	&	66	&	5.5	&	 \citet{croft2015measurement}	\\
\midrule
2003	&	65	&	31	&	 \citet{croft2015measurement}	\\
\midrule
2003	&	59	&	11	&	 \citet{croft2015measurement}	\\
\midrule
2004	&	78	&	3	&	 \citet{croft2015measurement}	\\
\midrule
2004	&	73	&	4.025	&	 \citet{croft2015measurement}
\\
\midrule
 2004	&	71	&	8	&	 \citet{croft2015measurement}	\\
\midrule
2004	&	71	&	7.1	&	 \citet{croft2015measurement}	\\
\midrule
2004	&	69	&	8	&	 \citet{croft2015measurement}	\\
\midrule
2004	&	67	&	24	&	 \citet{croft2015measurement}	\\
\midrule
2004	&	56	&	23	&	 \citet{croft2015measurement}	\\
\midrule
2005	&	73	&	6.4	&	 \citet{croft2015measurement}	\\
\midrule
2005	&	70	&	5	&	 \citet{croft2015measurement}	\\
\midrule
2005	&	66	&	12.5	&	 \citet{croft2015measurement}  \\
\midrule
2006	&	76.9	&	3.65	&	 \citet{croft2015measurement}
\\
\midrule
2006	&	74.92	&	2.28	&	 \citet{croft2015measurement}	\\
\midrule
2006	&	74	&	2	&	 \citet{croft2015measurement}	\\
\midrule
2006	&	74	&	6.3	&	 \citet{croft2015measurement}	\\
\midrule
2006	&	62.3	&	5.2	&	 \citet{croft2015measurement}
\\
\midrule
2007	&	76	&	8	&	 \citet{croft2015measurement}	\\
\midrule
2007	&	74	&	3.75	&	 \citet{croft2015measurement}
\\
\midrule
2007	&	68	&	10	&	 \citet{croft2015measurement}	\\
 \bottomrule
 \end{tabular}
 \end{table}

\begin{table}[H]\ContinuedFloat
\centering
\caption{{\em Cont.}}
\begin{tabular}{cccc}
\toprule
 \textbf{Date} & \boldmath{$H_0$} \textbf{(km s}\boldmath{$^{-1}$} \textbf{Mpc}\boldmath{$^{-1}$}\textbf{)} & \boldmath{$\pm$} & \textbf{Reference} \\ [0.5ex]
\midrule
2008	&	61.7	&	1.15	&	 \citet{croft2015measurement}
\\
\midrule
2009	&	74.2	&	3.6	&	 \citet{croft2015measurement}	\\
\midrule
2009	&	71	&	4	&	 \citet{croft2015measurement}	\\
\midrule
2009	&	70.5	&	1.3	&	 \citet{croft2015measurement}
\\
\midrule
2010	&	79.3	&	7.6	&	 \citet{croft2015measurement}	\\
\midrule
2010	&	69	&	11	&	 \citet{croft2015measurement}	\\
\midrule
2010	&	68.2	&	2.2	&	 \citet{croft2015measurement}	\\
\midrule
2010	&	66	&	5	&	 \citet{croft2015measurement}	\\
\midrule
  2011 & 73.8 & 2.4 &  \citet{riess20113} \\
 \midrule
 2011 & 74.8 & 3.1 &  \citet{riess20113} \\
 \midrule
 2011 & 74.4 & 6.25 &  \citet{riess20113} \\
 \midrule
 2011 & 68 & 5.5 &  \citet{chen2011median} \\
 \midrule
 2012 & 74.3 & 2.9 &  \citet{chavez2012determining}  \\
 \midrule
 2012 & 67 & 3.2 &  \citet{beutler20116df} \\
 \midrule
 2012 & 74.3 & 2.1 &  \citet{freedman2012carnegie} \\
 \midrule
 2013 & 68 & 4.8 &  \citet{braatz2012measuring} \\
 \midrule
 2013 & 68.9 & 7.1 & \citet{reid2013megamaser} \\
 \midrule
 2013 & 76 & 1.9 &  \citet{fiorentino2013cepheid} \\
 \midrule
 2014 & 69.6 & 0.7 &  \citet{bennett20141} \\
 \midrule
  2015 & 70.6 & 2.6 &  \citet{rigault2015confirmation} \\
 \midrule
 2015 & 68.11 & 0.86 &  \citet{cheng2015accurate} \\
 \midrule
 2016 & 73.24 & 1.74 &  \citet{riess20162} \\
 \midrule
 2017 & 68.3 & +2.7 $-$2.6 &  \citet{chen2017determining} \\
 \midrule
 2017 & 68.4 & +2.9 $-$3.3 &  \citet{chen2017determining} \\
 \midrule
 2017 & 65 & +6.5 $-$6.6 &  \citet{chen2017determining} \\
 \midrule
 2017 & 67.9 & 2.4 &  \citet{chen2017determining} \\
 \midrule
 2017 & 72.5 & +2.5 $-$8 &  \citet{bethapudi2017median} \\
 \midrule
 2017 & 69.3 & 4.2 &  \citet{braatz2017measurement} \\
 \midrule
 2018 & 66.98 & 1.18 &  \citet{addison2018elucidating} \\
 \midrule
 2018 & 64 & +9 $-$11 &  \citet{vega2018hubble} \\
 \midrule
 2018 & 73.48 & 1.66 &  \citet{riess2018new} \\
 \midrule
 2018 & 67 & 4 & \citet{yu2018hubble} \\
 \midrule
 2018 & 72.72 & 1.67 &  \citet{feeney2018clarifying} \\
 \midrule
 2018 & 73.15 & 1.78 &  \citet{feeney2018clarifying} \\
 \midrule
 2018 & 68.9 & +4.7 $-$4.6 &  \citet{hotokezaka2018hubble} \\
 \midrule
 2018 & 73.3 & 1.7 &  \citet{follin2018insensitivity} \\
 \midrule
  2018 & 67.4 & 0.5 &  \citet{chen2018two} \\
 \bottomrule
\end{tabular}
\end{table}

\begin{table}[H]\ContinuedFloat
\centering
\caption{{\em Cont.}}
\begin{tabular}{cccc}
\toprule
 \textbf{Date} & \boldmath{$H_0$} \textbf{(km s}\boldmath{$^{-1}$} \textbf{Mpc}\boldmath{$^{-1}$}\textbf{)} & \boldmath{$\pm$} & \textbf{Reference} \\ [0.5ex]
 \midrule
 2018 & 73.24 & 1.74 &  \citet{chen2018two} \\
 \midrule
 2019 & 67 & 3 &  \citet{kozmanyan2019deriving} \\
 \midrule
 2019 & 72.5 & +2.1 $-$2.3 &  \citet{birrer2019h0licow} \\
 \midrule
 2019 & 67.5 & +1.4 $-$1.5 &  \citet{dominguez2019new} \\
 \midrule
 2019 & 74.03 & 1.42 &  \citet{riess2019large} \\
 \midrule
 2019 & 67.8 & 1.3 & \citet{macaulay2019first} \\
 \bottomrule
\end{tabular}
\label{tab:a2}
\end{table}

%\end{center}



%% References
%%   \citet{key}          ==>>  [#]

\reftitle{References}

\end{document}